\pdfoutput=1
\documentclass[11pt]{article}
\usepackage{amsmath}
\usepackage[preprint]{acl}

\usepackage{times}
\usepackage{latexsym}
\usepackage{amssymb}
\usepackage{pifont}

\usepackage[T1]{fontenc}
\usepackage[utf8]{inputenc}

\usepackage{microtype}

\usepackage{inconsolata}

\usepackage{graphicx}
\usepackage{hyperref}
\usepackage{url}
\usepackage{booktabs}

\usepackage{multirow}
\usepackage[table]{xcolor}
\usepackage{arydshln}
\usepackage[skins]{tcolorbox} %必须标注skin，才能使用shadow命令显示阴影。
\tcbuselibrary{breakable}
\definecolor{lightblue}{RGB}{173, 216, 230}
\usepackage{soul}        % 提供 \hl 命令
\usepackage{xcolor}

\title{Words at Play: Benchmarking Audio Pun Understanding in Large Audio-Language Models}

\author{
 \textbf{Yuchen Su\textsuperscript{1}},
 \textbf{Shaoxin Zhong\textsuperscript{1}},
 \textbf{Yonghua Zhu\textsuperscript{2}\thanks{Corresponding author}},
 \textbf{Ruofan Wang\textsuperscript{1}},
 \textbf{Zijian Huang\textsuperscript{1}},
 \\
 \textbf{Qiqi Wang\textsuperscript{4}},
 \textbf{Na Zhao\textsuperscript{2}},
 \textbf{Diana Benavides-Prado\textsuperscript{3}},
 \textbf{Michael Witbrock\textsuperscript{1}}
\\
 \textsuperscript{1}School of Computer Science, University of Auckland, New Zealand
 \\
 \textsuperscript{2}Singapore University of Technology and Design, Singapore
 \\
 \textsuperscript{3}School of Electronic Engineering and Computer Science, Queen Mary University of London
 \\
 \textsuperscript{4}School of Statistics and Data Science, Nankai University, China
 \\
 \texttt{ysu132@aucklanduni.ac.nz, yonghua\_zhu@sutd.edu.sg}
 \\
}

\begin{document}
\maketitle
\begin{abstract}
Puns represent a typical linguistic phenomenon that exploits polysemy and phonetic ambiguity to generate humour, posing unique challenges for natural language understanding. Within pun research, audio plays a central role in human communication except text and images, while datasets and systematic resources for spoken puns remain scarce, leaving this crucial modality largely underexplored. In this paper, we present APUN-Bench, the first benchmark dedicated to evaluating large audio language models (LALMs) on audio pun understanding. Our benchmark contains 4,434 audio samples annotated across three stages: pun recognition, pun word location and pun meaning inference. We conduct a deep analysis of APUN-Bench by systematically evaluating 10 state-of-the-art LALMs, uncovering substantial performance gaps in recognizing, localizing, and interpreting audio puns. This analysis reveals key challenges, such as positional biases in audio pun location and error cases in meaning inference, offering actionable insights for advancing humour-aware audio intelligence. 
%We hope that APUN-Bench will inspire future research at the intersection of speech, humour, and advanced language understanding.
%APUN-Bench provides the foundation for future research at the intersection of speech, humour, and advanced language understanding.
\end{abstract}

\section{Introduction}

\begin{figure}[h] 
%\framebox[4.0in]{$\;$}
\resizebox{\columnwidth}{!}{
\includegraphics[width=\columnwidth,height=0.9\columnwidth]{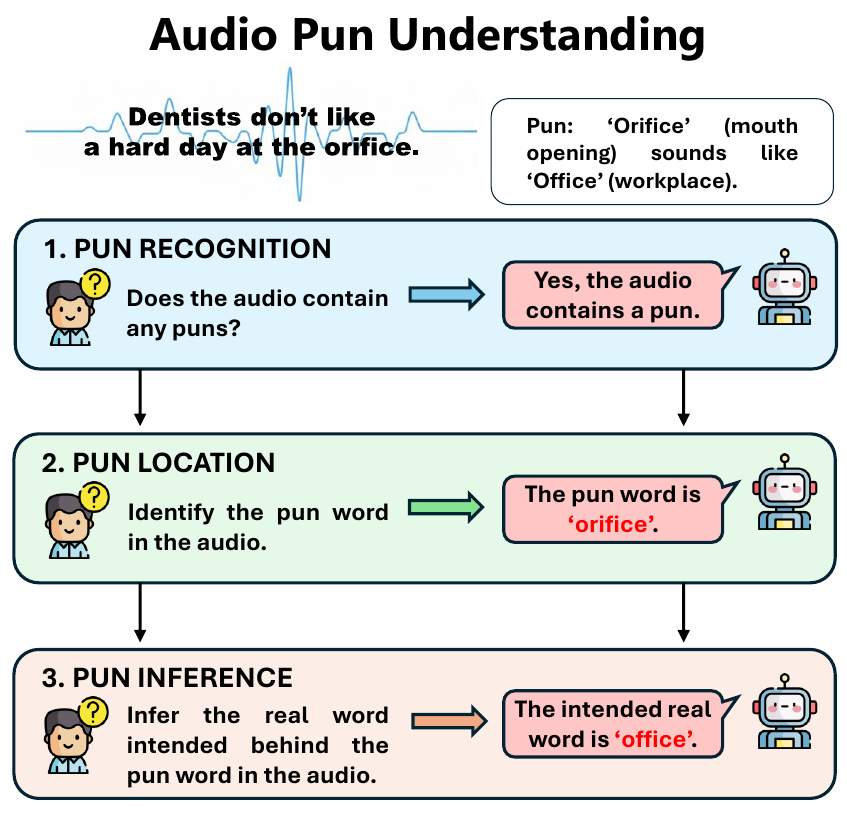}
}
\label{pun_explanation}
\caption{An example of the pun understanding in recognition, location and inference.}
\end{figure}

% Language is not only a medium of communication but also a crucial factor in shaping human–computer interaction and natural language generation. The ability to understand and process humor, particularly puns, can enhance the naturalness and approachability of intelligent systems, while also extending their value in applications such as education, voice assistants, and entertainment\cite{kumar2023natural}. As a typical linguistic phenomenon, puns exploit the polysemy of language and create cognitive tension through the complex interplay of phonetics, orthography, and generating humour \citep{partington2009linguistic}. Whether a machine can understand puns is not only an important indicator of its language analysis and reasoning capabilities, but also reflects its potential to process creative linguistic phenomena.\cite{kumar2023natural}
As a typical linguistic phenomenon, puns exploit the polysemy of language and create cognitive tension through the complex interplay of phonetics, orthography, and generating humour \citep{partington2009linguistic}. It is not only a vehicle for humorous expression, but also embodies a vital manifestation of human creativity and communicative flexibility \citep{aleksandrova2022pun}, with its inherent ambiguity makes it a rigorous test of true language understanding. \citet{miller2017semeval} and \citet{sun2022expunations} show that understanding pun phenomenon usually requires crossing three levels: determining whether there is a pun in the language fragment, locating the specific pun, and inferring the implied alternative words or double meanings. In terms of applications, puns play a significant role in areas such as education, voice assistants, and entertainment, thus they also provide a critical benchmark for evaluating the capabilities of general artificial intelligence \citep{morris2024position}.
% , making them a high-level benchmark for assessing the capabilities of artificial general intelligence (AGI). The AGI taxonomy proposed by \citet{morris2024position} emphasizes that achieving 90\% performance requires not only the successful completion of conventional tasks but also the ability to handle complex linguistic phenomena such as humour, metaphor, and puns. Among these, pun is not only a vehicle for humorous expression, but also embodies a vital manifestation of human creativity and communicative flexibility \citep{aleksandrova2022pun}. In general, understanding this phenomenon usually requires crossing three levels \citep{miller2017semeval,sun2022expunations}: determining whether there is a pun in the speech fragment, locating the specific pun, and inferring the implied alternative words or sub-level meanings. For intelligent systems, the ability to understand and process humour (especially puns) not only enhances fluency and credibility in natural interactions but also expands their value in applications such as education, voice assistants, and entertainment. 
% With the continuous advancement of artificial intelligence, the ability to understand these complex linguistic phenomena is becoming a new standard for measuring performace \citep{quan2025can}.

Audio puns play a particularly important role in interpersonal communication \citep{karpenko2017humor,palmann2025s}. Unlike text or images, audio is the most common medium of human communication, which carries not only linguistic content but also adds layers of depth and flexibility through acoustic cues. Moreover, audio puns are often shaped by phonetic ambiguity, pronunciation similarity, and acoustic variability \citep{benzeghiba2007automatic}, endowing them with rich expressive value in communication. It is worth noting that although some rare words may appear, their core challenge lies in speech-level ambiguity and contextual reinterpretation rather than vocabulary rarity \citep{zheng2022contextual}. Therefore, systematic investigation of audio puns contributes to a deeper understanding of human language and cognitive mechanisms, while presenting new challenges for the development of audio understanding and multimodal artificial intelligence.

Against this backdrop, benchmarking audio puns is critical for revealing the limitations of current LALMs and for improving their ability to process this complex linguistic phenomenon, while related research remains markedly underexplored. In recent years, researchers in natural language processing and computer vision have introduced several pun benchmarks \citep{ouyang2024punchbench,xu2024good,zhang2024creating} to systematically evaluate the recognition, explanation, and generation capabilities of large language models (LLMs) in text and image pun tasks, advancing the understanding of written and visual puns. However, systematic studies and publicly available datasets addressing puns at the phonetic level are still lacking. Meanwhile, existing state-of-the-art LALMs have focused primarily on relatively basic evaluation tasks, such as answering speech questions and multimodal understanding (speech–video) \citep{yang2024air,sakshi2024mmau,wang2024audiobench,gao2024benchmarking,zhang2025wildspeech,wang2025mmsu}. While these tasks are essential for testing fundamental audio comprehension, they do not specifically address more complex linguistic phenomena such as puns. This evaluation gap constrains the potential of LALMs to advance toward higher-level audio understanding.

In this work, we introduce APUN-Bench, a benchmark for audio pun understanding that comprises 4,434 audio samples, with complete annotations spanning three stages: pun recognition, pun word location, and pun meaning inference. The first two stages are identification tasks, requiring models to determine whether a given audio clip contains a pun and to further locate the punning words. Unlike previous research that has primarily focused on binary classification or word identification, our design additionally evaluates models’ ability to perform coarse-grained location in audio, thereby revealing their strengths and weaknesses at different levels. The third stage, pun meaning inference, assesses whether models can correctly infer the secondary meanings of three types of puns: heterographic, homographic, and homophonic. Additionally, we propose the pun dataset which integrates two parts: synthetic data and real-world speech data. The synthetic component is created by converting existing textual pun corpora into speech samples, while the real-world component is collected from publicly available speech resources to capture authentic speech contexts. We review and evaluate the quality of the speech data and corresponding annotations to ensure the overall reliability of the dataset. Our evaluation results on APUN-Bench reveal that current LALMs still have significant gaps in the recognition and understanding of audio puns. Overall, our key contributions are as follows:

\begin{itemize}
    \item We propose APUN-Bench, the benchmark for audio pun understanding, including 4,434-sample dataset combining synthetic and real-world speech with semi-automatic annotation and human verification.
    \item We comprehensively cover different stages of pun understanding, including pun recognition, pun word location, and pun meaning inference, as well as evaluate systematically on 10 open-source and proprietary LALMs using APUN-Bench.
    \item We provide an in-depth analysis of model performance, revealing key insights such as models’ positional bias in pun location and common error types in pun inference, which point to potential directions for future improvements.
\end{itemize}
%open-source and proprietary

\section{Related Work}
\subsection{Benchmarks for LALMs}
A growing number of benchmarks have been developed to evaluate LALMs. General-purpose resources such as Audiobench \citet{wang2024audiobench} and Airbench \citet{yang2024air} focus on recognition and generative comprehension across diverse acoustic conditions. More recent work extends this scope: \citet{sakshi2024mmau} and \citet{wang2025mmsu} assess multi-step spoken understanding and reasoning, while \citet{zhang2025wildspeech} highlights challenges in natural conversational settings. Other studies investigate specific dimensions, including cross-modal hallucination \citep{sung2024avhbench}, dialogue ambiguity \citep{gao2024benchmarking}, and reasoning strategies for structured inference \citep{ma2025audio,xie2025audioreasonerimprovingreasoningcapability}.

% Despite these advances, pun-related ambiguity in speech has not been systematically addressed. Existing benchmarks do not provide fine-grained localization of ambiguous segments, nor do they connect recognition, localization, and inference in a unified framework. Moreover, they lack explicit differentiation among homophonic, heterographic, and homographic puns, categories that are central to understanding phonological confusability and polysemy. To close this gap, we present \textbf{APUN-Bench}, the first benchmark dedicated to pun understanding in speech, structured around three stages: recognition, word-level localization, and semantic inference.

\begin{figure*}[h] 
\centering
%\framebox[4.0in]{$\;$}
\resizebox{2\columnwidth}{!}{
\includegraphics[width=2.5\columnwidth,height=0.9\columnwidth]{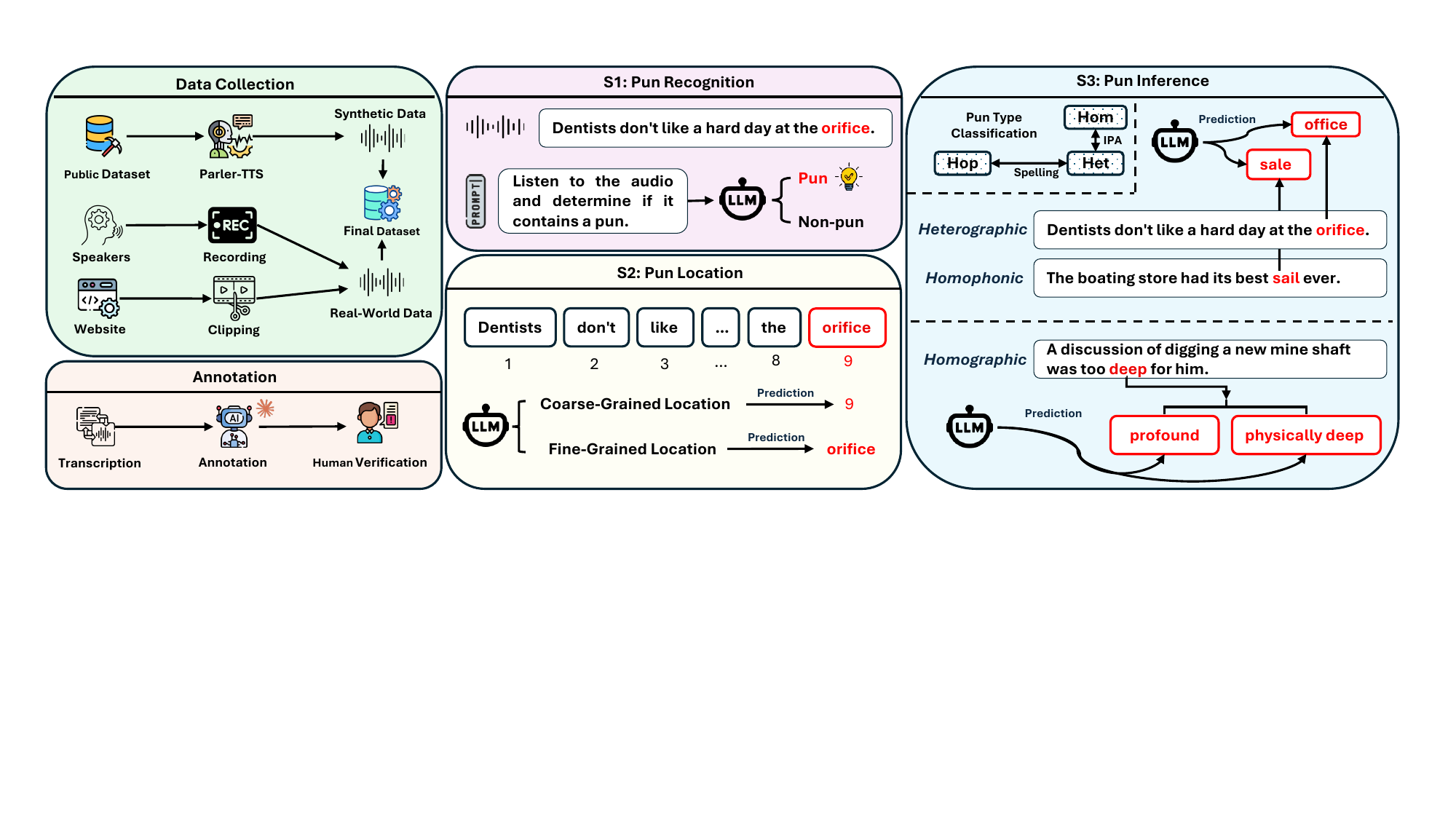}
}
\caption{The Overview of APUN-Benchmark, covering data construction, human annotation and three evaluation stages: (S1) Pun Recognition, detecting the presence of an audio pun; (S2) Pun Location, identifying the specific pun word in the audio; and (S3) Pun Inference, inferring its dual meanings.}
\label{overview}
\end{figure*}

\subsection{Pun Studies}
Puns, which rely on polysemy, homonymy and semantic conflict, are one of the most challenging linguistic phenomena for NLP models. Existing research on pun understanding has largely concentrated on recognition \citep{miller2017semeval,diao2018weca,zhou2020boating,jentzsch2023chatgpt}, location \citep{zou2019joint}, and explanation \citep{sun2022expunations}, but has also begun to explore multimodal extensions such as visual puns \citep{zhang2024creating}. Beyond understanding, scholars have further investigated automatic pun generation \citep{yu2018neural,luo2019pun,sun2022context,su2025survey} and systematic evaluation of the ability of large-scale language models to understand puns \citep{xu2024good}. However, as a central form of human communication, speech remains largely unexplored in pun research, and there is an urgent need for relevant benchmarks and systematic studies.
% 双关语作为一种依赖多义性、同音/同形或语义冲突的修辞手法，对于nlp模型来说最具有挑战性。作为双关语理解的任务之一，过去的研究不仅集中在文本双关语识别、双关语定位和双关语解释，还涉及到图像双关理解的研究。除此之外，还有一些双关语生成和评估大模型是否能够理解双关语上的研究。

\section{Benchmark}
Our goal is to evaluate the abilities of LALMs in understanding puns. Specifically, we aim to analyze the detection and deep semantic reasoning when processing puns in spoken language of current LALMs and provide insights for building more robust models with enhanced language understanding. As existing benchmarks focus primarily on pun research within textual modalities, we introduce the APUN-Bench, designed to measure a model’s ability to detect and infer pun-related phenomena in speech. Figure \ref{overview} provides an overview of the benchmark, including data collection, human annotation and task descriptions.

%We design three stages to comprehensively evaluate the ability of LALMs to understand puns. 
\textbf{Overview of APUN-Bench.} Inspired by pun research on unimodal text, APUN-Bench comprises three main stage categories: pun recognition, pun word location, and pun meaning inference. Pun recognition is a binary classification task to determine whether a speech segment contains a pun, aligning with prior work \cite{xu2024good}. The pun location stage focuses on the specific pun word, including localization and identification subtasks. The pun meaning inference stage evaluates understanding of the secondary meaning, with subtasks designed for homophonic, heterographic, and homographic puns \citep{su2025survey}. In this study, compound puns that involve multiple punning words within the same sentence are not considered. In total, APUN-Bench consists of 4,434 audio clips, including 2,971 audio puns paired with their corresponding pun words and inferred alternative words, as well as 1,463 negative samples.
% Inspired by pun research on unimodal text, APUN-Bench comprises three main stage categories: pun recognition, pun word location, and pun meaning inference. The pun recognition stage assesses the model’s ability to determine whether a given speech segment contains a pun, which is formulated as a binary (yes/no) classification problem, enabling intuitive and comparable evaluation, consistent with common practices in previous research \cite{xu2024good}. The pun location stage identifies the pun word in the audio, which is divided into two subtasks: pun word localization and pun word identification. Given a speech segment contained a pun, these subtasks evaluate the model’s ability to accurately locate and recognize the pun word. The pun meaning inference stage evaluates the model’s ability to infer the secondary meaning of a given pun. We categorize puns into three types: homophonic, heterographic and homographic \citep{su2025survey}, as well as design corresponding inference subtasks for each type. In this study, compound puns that involve multiple punning words within the same sentence are not considered. Overall, APUN-Bench consists of 4,434 audio clips, including 2,971 audio puns paired with their corresponding pun words and inferred alternative words, as well as 1,152 negative samples.

\subsection{Data Collection}
The dataset construction consists of two main components: (1) Synthetic Audio Dataset: Audio samples synthesized from existing pun text data; (2) Real-world Audio Dataset: Audio data collected from real-world sources, including human recordings and publicly available websites, followed by editing and annotation.
%在数据集的构建主要分成了两个部分：（1）合成语音数据集：通过目前已有的双关语文本数据合成语音信息（2）公开数据集：从公开网站中收集语音数据并进行剪辑和标注

\textbf{Synthetic Audio Dataset.}
One challenge in constructing the benchmark is collecting audio data and performing fine-grained annotations. To address this, we used the existing SemEval 2017 pun text dataset \citep{miller2017semeval}, which comprises 2,389 positive instances (puns) and 1,152 negative instances (non-puns). The positive set comprises instances annotated with both the punning word and its corresponding replacement. Then, we employ state-of-the-art speech synthesis Parler-TTS \citep{lacombe-etal-2024-parler-tts,lyth2024natural} to randomly assign each text to one of 34 distinct synthetic speakers, ensuring diversity in speaker characteristics. More details can be found at Appendix \ref{Synthesis Process}.
%The positive set contains 1,291 homographic puns and 1,098 homophonic or heterographic puns, each annotated with the punning word and its corresponding replacement.

\textbf{Real-world Audio Dataset.} In addition to the synthetic audio dataset, we collect publicly available videos and manually segment the speech content from the \textit{O. Henry Museum Pun-Off} \footnote{https://www.punoffatx.brushsquaremuseums.org/}, a long-standing global pun competition where contestants must create puns within a limited time to compete for the championship. To address the limited availability of public audio data, we further collect data from pun websites \footnote{https://pun.me/} and create realistic pun speech by manual recording. Some details can be found in Appendix \ref{human annotation}. Furthermore, we constructed negative samples from real speech. The dataset contains 582 positive samples and 311 negative samples. These annotated data effectively introduce the challenge of understanding puns in audio, requiring models to accurately detect and interpret them under complex conditions. In total, our dataset comprises 3,541 synthetic and 892 real-world audio samples, ensuring diversity across data sources. Table \ref{Statistics} presents the statistics of both synthetic and real-world data.
%为了弥补公众语音数据的不足，我们还收集了双关语网站中的数据并人为录制了真实双关语语音。表x对合成数据和真实数据进行了数据统计

\begin{table}[t]
\resizebox{\columnwidth}{!}{
\centering
\begin{tabular}{lccccc}
\toprule
\textbf{Dataset} & \textbf{Heterographic} & \textbf{Homophonic} & \textbf{Homographic} & \textbf{Negative} & \textbf{Overall} \\
\midrule

Synthetic Dataset  &794 &304 &1298 & 1,152 & 3,541 \\
\hspace{1em} - Segment Clip (s)  & 4.18 & 4.36 & 3.88 & 3.47 & 3.97\\
\hspace{1em} - Sentence Length  & 12.21 & 13.89 & 11.68 & 8.50 & 11.57\\
Real-World Dataset      & 223 & 72&287& 311 &582\\
\hspace{1em} - Segment Clip (s)  & 6.36 & 6.16 & 5.34 & 6.73 & 6.14\\
\hspace{1em} - Sentence Length  & 11.12 & 11.40 & 11.55 & 10.88 & 11.23 \\
\bottomrule
\end{tabular}
}
\caption{APUN-Bench data statistics, including average audio clip duration and sentence length.}
\label{Statistics}
\end{table}

\subsection{Dataset Construction}
\label{task definition}
We provide detailed descriptions for 3 different stages, comprising a total of 6 subtasks.

\textbf{Stage 1: Pun Recognition.} This task assesses the model’s ability to detect the presence of puns in audio, which serves as a prerequisite for more fine-grained comprehension and reasoning tasks. Inspired by \cite{xu2024good}, we construct the question format that include both pun and non-pun to mitigate potential biases, whose primary question format is "Determine whether the given audio is a pun/non-pun".

\textbf{Stage 2: Pun Location.} This task evaluates the ability of LALMs to localize pun words within spoken sentences. We investigate whether a model that can already detect the presence of a pun in audio may still struggle to accurately identify the specific pun word. To enable more granular evaluation, we divide this task into two levels: (1). Coarse-grained Location: The model is evaluated for its ability to accurately locate the approximate location of pun words in the audio, without requiring the semantic accuracy of the identified words. Even if the model locates a semantically inaccurate word, as long as its location makes sense, it is considered correct. (2). Fine-Grained Location: this subtask further requires the model to accurately identify actual puns word in the audio.

For real-world speech data, manually labeling each audio segment in fine-grained detail is time-consuming. Previous research has demonstrated that leveraging LLMs to assist in dataset construction is both efficient and feasible \citep{gong2023listen,liu2023llava,hyun-etal-2024-smile}. \citet{xu2024good}) has demonstrated that Claude ranks among the best existing models for understanding text-based puns. Accordingly, we first transcribe the audio into text using the Whisper-large model \citep{radford2022whisper}, and then employ Claude-Opus (particularly Claude-Opus-4-1) \citep{claude_systemcard_2025} to identify and locate the puns. To enhance its task comprehension and annotation accuracy, we prompt Claude-Opus with a small set of manually annotated examples along with feedback. More details can be found in the appendix \ref{Annotation Prompts}.

\textbf{Stage 3: Pun Inference.} This task evaluates the ability of LALMs to infer the intended meaning of puns in spoken sentences. Based on the characteristics, puns are classified into three categories for evaluation: heterographic, homophonic, and homographic: (1) Heterographic puns involve words with similar but not identical pronunciations. For this type, the model needs to infer the correct pronunciation and the corresponding intended meaning substitute based on the given pun; (2) Homophonic puns involve words with identical pronunciations but different spellings. It is similar to heterographic puns, where the task is to identify the intended alternative word. (3) Homographic puns involve words with the same written form but different meanings. The model is required to generate phrases that capture both meanings and compare them with the gold standard, thereby assessing its ability to understand polysemy.

Based on the characteristics of different pun categories, we adopt a differentiated classification strategy. For homographic puns, which rely on lexical polysemy without spelling differences, we classify them by checking the consistency of word spellings. For homophonic and heterographic puns, whose defining feature lies in the phonetic differences between the punning word and its substitute, we transcribe the target words into International Phonetic Alphabet (IPA) symbols and determine their category based on whether their pronunciations match.

% For the synthetic dataset, since the original data does not inherently distinguish between homophonic and heterographic puns, we leverage its predefined characteristics (see Section \ref{task definition}) and transcribe the words into International Phonetic Alphabet (IPA) symbols to enable category differentiation automatically. For the real-world data set, classification is performed manually based on linguistic analysis.
%基于双关语不同类别的特性，对于homographic来说，因为其集中于词语的一词多义而不区分单词的拼写，因此我们匹配单词拼写的一致性来进行分类。对于homophonic and heterographic来说，由于其特性是区别于双关语词和替换词发音特征，我们将单词转录为国际音标（IPA）符号，并通过判断其IPA是否一致来进行区分

To infer replacement words for heterographic and homophonic puns, we employed a similar process to Stage 2: given a pun, Claude-Opus is prompted to infer and generate suitable replacements. It is worth noting that for homographic puns, which involve polysemy, we provide Claude-Opus with targeted prompts to generate two distinct sets of phrases representing different meanings of the pun. All outputs are subsequently verified and manually corrected to mitigate potential model bias and produce the final gold-standard annotations.

% \subsection{Dataset Construction}
% \label{Dataset construction}

% \textbf{Stage 1: Pun Recognition.}
% % 我们在该阶段利用数据收集阶段构建正负样本进行实验，负样本被标记为普通句子以对双关语进行区分，进行二分类
% In this stage, positive and negative samples are constructed from the collected data to perform the binary classification task. We label negative samples as ordinary sentences to distinguish them from puns.

% \textbf{Stage 2: Pun Word Location.} The goal of this stage is to extract puns from speech information and mark their specific locations in the audio sentences. 

% \textbf{Stage 3: Pun Inference.} After identifying the pun words, we infer the vocabulary corresponding to their secondary meanings and classify the pun data into homographic, homophonic and heterographic categories. 

\textbf{Human Annotation.} During benchmark construction, we conduct multiple rounds of manual development and reviews. In the raw data collection phase, each audio clip is rigorously examined each audio clip for pun presence, pun words, categories, and replacement words to ensure the quality of the audio pun data. Audio samples with substandard quality are re-collected to ensure data reliability. In addition, when inferring double meanings for homographic puns in the synthetic speech dataset, we manually review the generated polysemous outputs, revise inappropriate content, and correct errors to obtain gold-standard annotations. The manual error correction rate is approximately 6\%, and all checks are performed by the authors. Furthermore, a final human verification process is conducted to ensure the overall reliability of the dataset. More details can be found at Appendix \ref{Human Verification}

\section{Evaluation}
\label{evaluation}

% \subsection{Experimental Settings}
\textbf{Baseline Models.} We evaluate APUN-Bench on a range of 10 recent LALMs, widely cited in the literature and capable of jointly processing text and audio information, including Qwen2-Audio-Instruct \citep{Qwen2-Audio}, Audio-Reasoner \citep{xie2025audioreasonerimprovingreasoningcapability}, Audio Flamingo 3 \citep{goel2025audio}, Qwen-2.5-omni \citep{Qwen2.5-Omni}, MiniCPM \citep{yao2024minicpm}, Gemini 2.0 Flash \citep{jaech2024openai}, SALMONN \citep{tang2023salmonn}, GPT4o-Audio \citep{hurst2024gpt}, Omni-R1 \citep{rouditchenko2025omni} and MERaLiON2 \citep{he2024meralionaudiollmtechnicalreport,huang2025meraliontextllmcrosslingualunderstandinglarge}. Furthermore, we also evaluate cascaded systems, including combinations of SenseVoiceSmall \citep{an2024funaudiollm} with GPT4o and Gemini 2.0 Flash, as well as Whisper-Large \citep{radford2022whisper} paired with GPT4o and Gemini 2.0 Flash. Each model is configured with a temperature of 0 to ensure deterministic outputs.

%Accuracy represents the overall proportion of correct classifications, while precision and recall indicate the model’s performance in identifying “pun” and “non-pun” sentences, respectively.

\textbf{Evaluation Strategy.} We adopt appropriate evaluation metrics for different tasks. For pun recognition tasks, we use accuracy, precision, recall, and F1 score as evaluation criteria, following the common practice in previous pun recognition research \citep{gepalova2024clef,xu2024good}.  Inspired by \citep{sung2024avhbench}, we also report the proportion of “pun” responses produced by each model to examine potential biases in pun recognition.

For the pun word location stage, the coarse-grained setting requires only identifying the approximate location of the pun within the speech. Specifically, we first transcribe the target audio and obtain the IPA representation of each word, then compute the character-level edit distance between these representations and the IPA of the pun word predicted by the model. In addition, we incorporate phonetic characteristics by defining sets of vowels and consonants, mapping phonemes into feature vectors, and calculating their cosine similarity. We then combine these two similarity measures to select the word position with the highest overall similarity as the prediction result, as shown in the formula below:
\begin{equation}
S_{\text{edit}}(w_i) = 1 - \frac{d_{\text{edit}}(\phi(w_i), \phi(\hat{w}))}
{\max(|\phi(w_i)|, |\phi(\hat{w})|)},
\end{equation}
\begin{equation}
S_{\text{cos}}(w_i) = \cos\big(v(\phi(w_i)), v(\phi(\hat{w}))\big),
\end{equation}
\begin{equation}
S(w_i) = S_{\text{edit}}(w_i) + S_{\text{cos}}(w_i),
\end{equation}
%(1-\alpha) \,  , and \(\alpha \in [0,1]\) is a weighting factor

where \(\phi(\cdot)\) denotes the IPA representation of a word,  \(d_{\text{edit}}\) is the Levenshtein distance, and \(v(\cdot)\) maps phonemes into feature vectors. For fine-grained location, an exact match is required to identify puns within speech. We first perform lemmatization on the words and then directly match them with the pun words predicted by the model to obtain the final evaluation outcome.

During the pun inference phase, results for heterographic and homophonic puns are evaluated by exact matching between the model-predicted pun words and the ground-truth annotations, as illustrated in the formula below:

\begin{equation}
\text{Eval}(p) = 
\begin{cases}
1, & \hat{w} = w^{*}, \\
0, & \text{otherwise},
\end{cases}
\end{equation}

where $\hat{w}$ is the model-predicted pun word and $w^{*}$ is the ground-truth alternative words. For homographic puns, as described in Section \ref{task definition}, we first construct a gold standard capturing the dual meanings of each pun. The model’s predicted meanings are then compared with this gold standard using similarity matching, and the two pairs with the highest similarity are selected. The formula is illustrated as below:

\begin{equation}
\resizebox{0.9\hsize}{!}{$
\text{Eval}(p) = 
\begin{cases}
1, & S(\hat{m}_{j_1}, m^{*}_{i_1}) > \tau \;\land\; S(\hat{m}_{j_2}, m^{*}_{i_2}) > \tau, \\
0, & \text{otherwise}.
\end{cases}
$}
\end{equation}

where $m^{*}_i$ are the gold-standard meanings of a homographic pun, 
$\hat{m}_j$ are the model-predicted meanings, 
$S(\cdot,\cdot)$ denotes the similarity function, 
and $\tau$ is the threshold. Based on empirical experiments, pairs with similarity scores above 0.3 ($\tau$ = 0.3) are retained and regarded as positive examples for this task.

\begin{table*}[t]
\resizebox{\textwidth}{!}{
\centering
\begin{tabular}{lccccccccccc}
\toprule
\multirow{2}{*}{\textbf{Model}} & \multirow{2}{*}{\textbf{Size}} & \multicolumn{5}{c}{\textbf{Pun Recognition}} & \multicolumn{2}{c}{\textbf{Pun Location}} & \multicolumn{3}{c}{\textbf{Pun Inference}} \\
\cmidrule(lr){3-7} \cmidrule(lr){8-9} \cmidrule(lr){10-12}
& & Acc. ($\uparrow$) & Pre ($\uparrow$) & Rec ($\uparrow$) & F1 ($\uparrow$) & Pun ($\%$) 
& Coa ($\uparrow$) & Fin ($\uparrow$) & Heg ($\uparrow$) & Hog ($\uparrow$) & Hop ($\uparrow$)\\
\midrule
% Dendrite     & & &  &  &  &  &  &  &  \\
\addlinespace[0.5ex]   % 给虚线下方加空隙
\multicolumn{12}{c}{\textit{Synthetic Dataset}} \\ \hdashline
\addlinespace[0.8ex]   % 给虚线下方加空隙
Qwen2-Audio-Instruct & 7B &45.78 &71.70 &39.45 &50.90 &38.06&50.81&31.77&13.47 & 17.91 & 26.97 \\
SALMONN & 13B & 51.91 & 72.51 & 52.60 & 60.97 & 51.81 &34.41 &22.35 & 15.23 & 24.63 & 25.32\\
Audio-Reasoner &7B & 55.27& 72.62 & 59.81 & 65.59 & 54.47& 50.39 & 32.31 & 23.05 & 27.25&37.50\\
Audio-Flamingo-3 & 7B&37.51&59.25&	40.02	&47.78&	48.24&	36.37&	27.83&	33.25&	29.35&	56.91 \\
% R1-AQA & 7B & 54.66 &70.64 & 62.47 & 66.31 & N/A\\
% Qwen-2.5-omni & 3B \\
Qwen-2.5-omni     &7B& 64.04 & 78.71 & 68.07 & 73.00 &61.76 & 60.98 & 48.01 & 40.55 & 32.38 & 60.19 \\
MiniCPM &8.7B&62.31&78.00&65.77&71.37 &27.92 &59.94 & 44.62 & 39.54 & 37.36&67.43\\
%LLaSO-Base & 3.8B \\
Omni-R1 & 8.9B & 65.28 & 77.39 & 72.58 & 74.91 & 66.97 & 63.49 & 49.43 & 44.20 & 39.73 & 63.48\\
% GLM-4-Voice \\
MERaLiON2 & 10B &73.40 &75.66 &92.56 &83.26 & \textbf{87.37} & 60.44 & 43.78 & 50.88 &29.43 & 79.93\\
% \midrule
% \addlinespace[0.5ex]   % 给虚线下方加空隙
% \multicolumn{12}{c}{\textit{Closed-Source LALMs}} \\ \hdashline
% \addlinespace[0.8ex]   % 给虚线下方加空隙
Gemini 2.0 Flash \ding{79} & - & \textbf{80.00}& \textbf{91.04} & 79.86 &85.08&62.39&65.80&48.34&52.39&53.29&92.11\\
% Gemini 2.5 Flash & - &  \\
GPT4o-Audio \ding{79} & - &78.04& 78.48 & \textbf{95.92} & \textbf{86.33} & 83.87 & 75.97 & 56.63 & 59.06 &57.81 & 85.19 \\
% \addlinespace[0.5ex]   % 给虚线下方加空隙 
% \hdashline \multicolumn{12}{c}{\textbf{\textit{Cascaded Models}}} \\
\addlinespace[0.5ex]   % 给虚线下方加空隙
\hdashline
\addlinespace[0.5ex]   % 给虚线下方加空隙
SenseVoiceSmall+GPT4o & - & 79.44 & 84.52 & 87.16 & 85.82 & 73.57 & 75.04 & 56.05 & \textbf{67.50} & 67.79 & 98.02\\
SenseVoiceSmall+Gemini2.0-Flash & - & 78.65 & 84.12 & 86.64 & 85.36 & 73.47 & 74.08 & 55.46 & 55.16 & 60.10 & 78.28\\
Whisper-Large+GPT4o & - & 80.29 & 87.28 & 84.69 & 85.97 & 68.91 & \textbf{79.19} & \textbf{62.78} & 67.00 & \textbf{70.49} & \textbf{98.35} \\
Whisper-Large+Gemini2.0-Flash & - & 77.47 & 84.37 & 84.58 & 84.47 & 71.46 & 77.56 & 60.86 & 55.54 & 62.94 & 89.80 \\
\midrule
\addlinespace[0.5ex]   % 给虚线下方加空隙
\multicolumn{12}{c}{\textit{Real-world Dataset}} \\ \hdashline
\addlinespace[0.8ex]   % 给虚线下方加空隙
% \multicolumn{12}{c}{\textbf{\textit{End-to-End Models}}} \\
Qwen2-Audio-Instruct & 7B & 58.64 & 95.11 & 37.61 & 53.90 & 25.53 & 49.05 & 24.95 & 23.87 & 21.59 & 26.02 \\
SALMONN & 13B & 44.86 & 57.07 & 60.41 & 58.69 & 68.56 & 28.82 & 14.11 & 18.38 & 23.15 & 16.43 \\
Audio-Reasoner & 7B & 70.99 & 89.97 & 61.94 & 73.37 & 41.14 & 48.71 & 25.98 & 34.97 & 25.70 & 38.35 \\
% Qwen-2.5-omni & 3B &  &  &  &  &  &  &  &  &  &  \\
Audio-Flamingo-3&7B&63.65&97.39&45.09&61.64&29.99&45.07&31.78&37.66&25.96&50.68 \\
Qwen-2.5-omni & 7B & 71.68 & \textbf{99.09} & 56.79 & 72.21 & 37.12 & 59.21 & 42.85 & 50.00 & 35.78 & 57.53 \\
MiniCPM & 8.7B & 74.35 & 98.08 & 61.62 & 75.68 & 40.80 & 54.73 & 38.03 & 43.69 & 34.86 & 54.79 \\
% LLaSO-Base & 3.8B &  &  &  &  &  &  &  &  &  &  \\
Omni-R1 & 8.9B & 83.50 & 97.37 & 76.59 & 85.74 & 50.95 & 58.86 & 41.31 & 44.59 & 35.00 & 57.53 \\
MERaLiON2 & 10B & \textbf{88.18} & 87.05 & 96.04 & \textbf{91.32} & \textbf{71.46} & 54.56 & 35.45 & 49.54 & 32.28 & 62.50 \\
Gemini 2.0 Flash \ding{79} & - & 87.51 & 84.82 & \textbf{98.51} & 91.15 & 60.98 & 67.52 & 51.03 & 62.78 & 47.36 &  82.85  \\
% Gemini 2.5 Flash & - &  \\
% Gemini 2.5 Flash & - &  &  &  &  &  &  &  &  &  &  \\
GPT4o-Audio \ding{79} & - & 85.36 & 87.02 & 92.56 & 89.71 & 66.89 & 66.95 & 50.08 & 60.98 & 50.70 & 76.71 \\ 
% \addlinespace[0.5ex]   % 给虚线下方加空隙 
% \hdashline \multicolumn{12}{c}{\textbf{\textit{Cascaded Models}}} \\
\addlinespace[0.5ex]   % 给虚线下方加空隙
\hdashline
\addlinespace[0.5ex]   % 给虚线下方加空隙
SenseVoiceSmall+GPT4o & - & 84.94&	96.45&	79.69&	87.27&	53.51&	67.29&	43.78&	71.74&	\textbf{67.36} & \textbf{87.67}\\
SenseVoiceSmall+Gemini2.0-Flash & - & 83.61	&92.56	&81.41	&86.63	&56.97	&67.12	&44.06&	50.06&	49.47	&86.30 \\
Whisper-Large+GPT4o & - & 85.20	&98.26	&78.47&	87.25&	51.28	&\textbf{76.41}	&\textbf{55.76}&	\textbf{72.19}&	67.01&	84.93\\
Whisper-Large+Gemini2.0-Flash & - & 83.08	&95.34	&77.89	&85.74	&52.73&	69.53	&49.74	&52.46	&52.98	&82.19\\
% \addlinespace[0.5ex]   % 给虚线下方加空隙
% \hdashline
% \addlinespace[0.5ex]   % 给虚线下方加空隙
% Human (Sample) & - & \\

\bottomrule

\end{tabular}
}
\caption{Evaluation Results of the Synthetic and Real-World Datasets on APUN-Bench in various LALMs and cascaded systems. \ding{79} represents the proprietary LALMs and the symbol '–' indicates that the model size is not publicly disclosed. Coa and Fin repesent the accuracy of coarse-grained and fine-grained locations. Hog, hop and heg denote the accuracy of homographic puns, homophonic puns and heterographic puns.}
\label{main}
\end{table*}

\section{Results and Discussion}
Based on our proposed APUN-Bench, we conduct a systematic evaluation of LALMs in pun understanding. We highlight several key findings from the experimental results, analyze the potential underlying causes of the observed phenomena and explore promising directions for improvement.

\subsection{Main Results}

We systematically evaluate baseline models across the three stages of APUN-Bench on both synthetic and real-world datasets, with the results summarized in Table \ref{main}. In the pun recognition stage, most baseline models perform above chance level (50\% accuracy). However, with the exception of GPT4o-Audio, Gemini2.0-Flash, and MERaLiON2, their recall rates are generally low, suggesting that the models tend to be overly conservative in identifying puns and are prone to false negatives. In the pun location stage, models are able to identify the approximate positions of more than half of the puns in the coarse-grained location task. Nevertheless, their performance drops substantially in the fine-grained detection task when recovering the exact pun words, revealing LALMs’ limited ability to accurately transcribe specific pun words. In the pun inference stage, models achieve markedly better results on homophonic puns than on heterographic puns, consistent with findings that homophony constitutes only about 2–3\% of the English lexicon \citep{marian2012clearpond}, whereas a four-letter English word has on average about 10.33 phonological neighbors.

Moreover, proprietary models consistently outperformed open-source speech models in APUN-Bench, especially in pun location and inference, underscoring the gap between the two in deep semantic understanding. We further conduct McNemar’s test to examine whether the results are statistically significant. For example, GPT4o-Audio achieves a highly significant advantage over MiniCPM in fine-grained pun location ($p < 1 \times 10^{-10}$). Overall, although closed-source models exhibit certain advantages across tasks, current auditory language models still face significant challenges in detecting and understanding puns, with overall performance remaining unsatisfactory.

Furthermore, we report the performance of cascaded models on audio pun understanding. Overall, cascaded systems underperform LALMs on pun recognition, but outperform them on pun word location and pun meaning inference. This demonstrates that in word-level tasks, the powerful ASR model can reduce transcription errors, allowing cascaded systems to better locate and inference pun words. In addition, GPT-4o achieves better performance than other text-based model, highlighting the ability of strong downstream reasoning. The results of word error rate can be found at Appendix \ref{Word Error Rate section}.
%For pun word location, models using Whisper-Large consistently outperform those based on SenseVoice-Small, indicating more accurate transcriptions. For pun meaning inference, when using the same ASR model, GPT-4o achieves better performance than other text-based models, highlighting the importance of strong downstream reasoning.

%
% \subsection{Pun Recognition Preferences}

\begin{figure}[h] 
%\framebox[4.0in]{$\;$}
\resizebox{\columnwidth}{!}{
\includegraphics[width=\columnwidth,height=0.7\columnwidth]{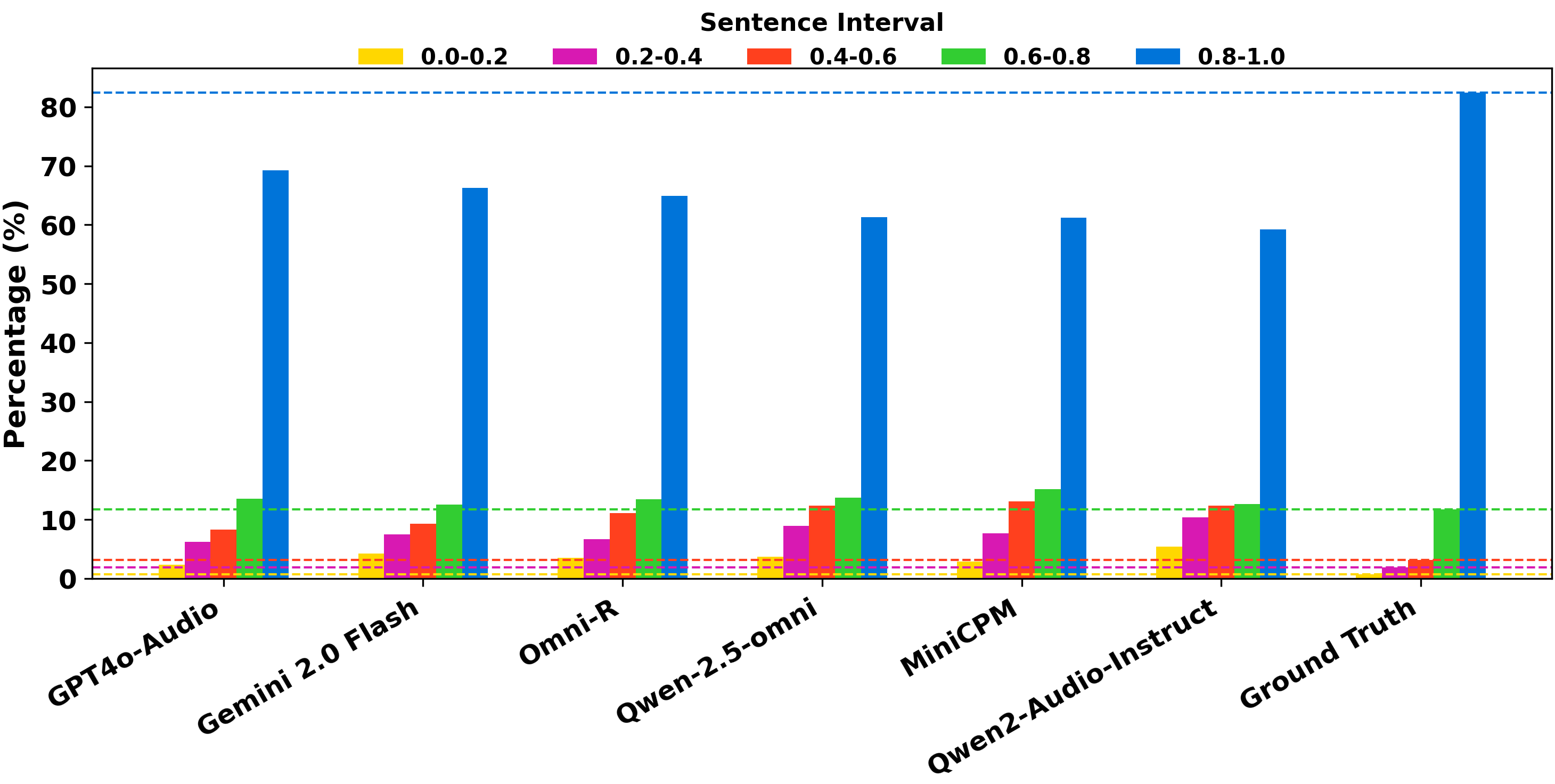}
}
\caption{Statistical comparison of predicted in different LALMs and ground-truth pun positions in APUN-Bench, showing the distribution of pun words across different sentence intervals (beginning, middle, end of the sentence). }
\label{pun_location}
\end{figure}

\begin{figure*}[h] 
\centering
%\framebox[4.0in]{$\;$}
\includegraphics[width=0.9\textwidth]{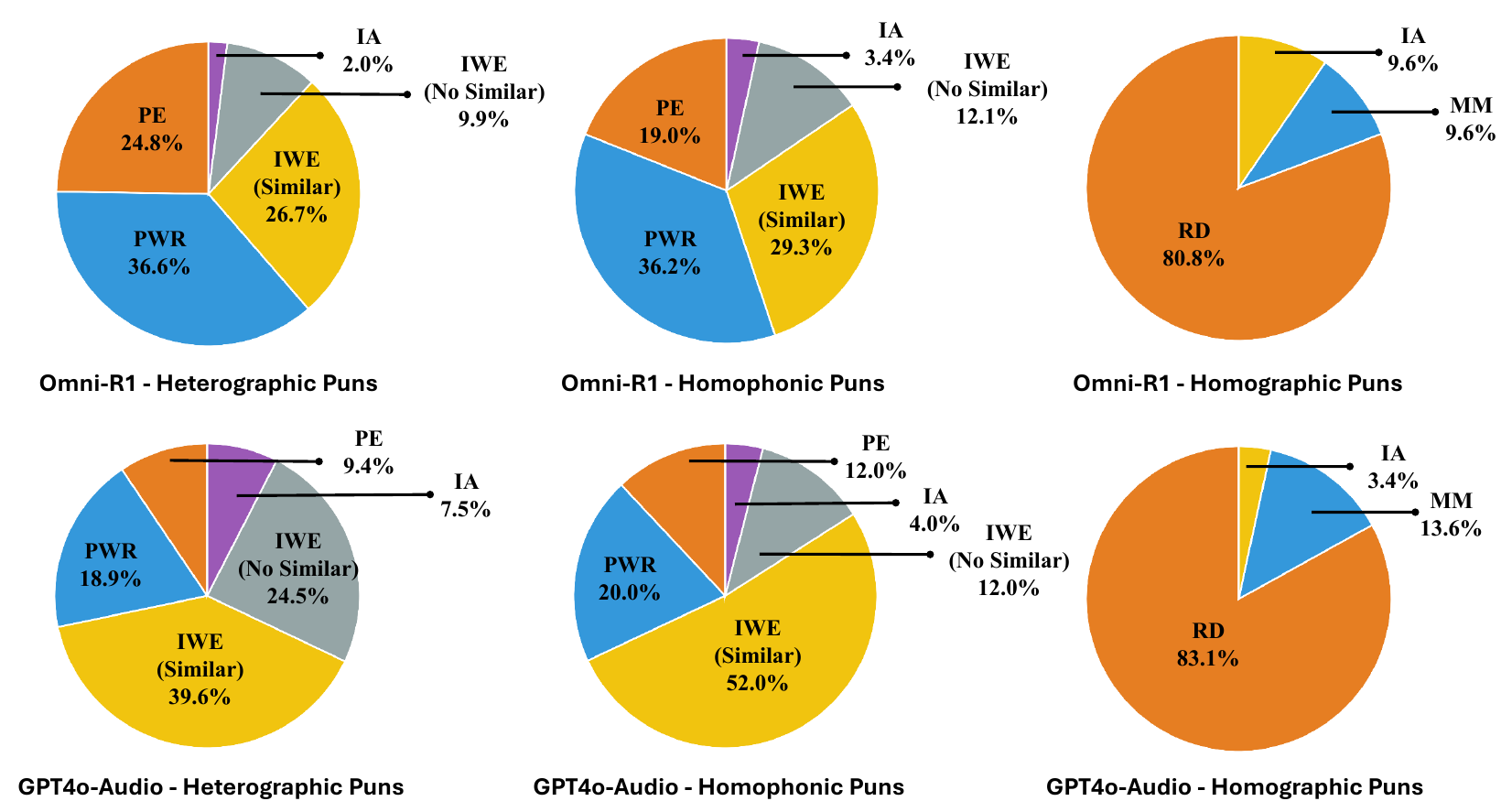}
\caption{The distribution of error types in pun inference for Omni-R1 and GPT4o-Audio across heterographic, homographic, and homophonic puns.}
\label{pun_inference_error_results}
\end{figure*}

\subsection{Analyzing the Position of Pun Words in Sentences}
According to the incongruity theory of humour, punchlines typically appear at the end of a sentence \citep{shahaf2015inside}. Building on this, we analyze the positional preferences of different models in identifying puns and compared them with manual annotation results, as shown in Figure \ref{pun_location}. Sentence Interval refers to the normalized sentence position; for example, 0–0.2 denotes the first 20\% of the whole sentence. The annotation data indicate that the vast majority of puns occur within the last 40\% of a sentence, with fewer than 10\% located in the first 60\%, further supporting the incongruity theory. In contrast, most model predictions exhibit a more uniform distribution: although the last 20\% still contains the highest proportion of predicted puns, more than 20\% appear in the first 60\%. These findings suggest that guiding models to focus more on the latter half of a sentence may improve their performance in pun location. 

To examine this hypothesis, we conduct a simple experiment. When designing prompts for pun location, we introduce the information that strengthens the model’s attention to the latter half of the sentence. The following shows an improved segment of the prompt:
%例如，下面是部分prompt中的改进。我们在设计模型定位双关语时的prompt中注入增强句子后半部分注意力的信息。
\begin{tcolorbox}[breakable, title=Improved Segment of Prompt]
[...]
Identify the pun word in the audio, \sethlcolor{lightblue}\hl{paying particular attention to the latter half of the sentence.}
[...]
\end{tcolorbox}
We conduct evaluations on Qwen-2.5-omni-7B and MiniCPM, achieving 63.49\% and 62.32\% accuracy in coarse-grained pun location, respectively, surpassing the baseline models by 2.51\% and 2.38\% percentage points, as shown in Table \ref{main}. For fine-grained location, the models achieve 49.56\% and 46.42\% accuracy, outperforming the baselines by 1.55\% and 1.80\% percentage points, respectively. While the overall improvements are limited, the results indicate that emphasizing the latter part of a sentence may contribute to better pun localization and provide initial insights for future research.
%These experimental results demonstrate that emphasizing the latter half of a sentence enhances model performance in pun localization, offering valuable insights for future research on pun location tasks.
%我们分别在qwenomni和MiniCPM上进行了测试，在粗粒度pun location中分别为63.49%和62.32，分别超过baseline模型2.51和2.38。而在细粒度检测的准确率分别为49.56和46.42，分别超过baseline模型1.55和1.8。这些实验结果表明，模型在关注句子后半部分权重时对双关语词定位的性能会提高，这为后续的双关语定位任务请提供了启发式思路。

% \subsection{Understanding}

\subsection{Audio Pun Inference: Where do LALMs Fall Short?}

Figure \ref{pun_inference_error_results} presents the distribution of error types for Omni-R1 and GPT4o-Audio across 450 instances, with 150 examples for each pun category. Detailed definitions of error types are provided in Table \ref{pun_inference_error}. For heterographic and homophonic puns, the dominant error type in both models is Word Inference Error (including Similar and No Similar), accounting for more than one-third of the errors. Similarly, for homographic puns, the primary error type shifts to Reasoning Disorder, which constitutes over three-quarters of the errors, while Meaning Mismatch accounts for only about 10\%. This indicates that although the models are generally capable of perceiving speech and identifying one layer of pun meaning, they encounter substantial difficulties in uncovering and restoring the secondary meaning. Moreover, for heterographic and homophonic puns, GPT4o-Audio exhibits a markedly lower proportion of Pun Word Repeat errors compared to Omni-R1, suggesting that larger-scale models are less prone to rigidly reproducing the pun word itself as output. Finally, Irrelevant Answer represents the smallest proportion of errors across all categories and both models, implying that most models are able to understand the task requirements and attempt to provide answers related to pun reasoning rather than completely unrelated outputs. Overall, our error analysis suggests that enhancing speech comprehension is crucial for improving performance. Building on prior research, this can be achieved through the targeted design of audio chain-of-thought \cite{ma2025audio} reasoning mechanisms and by expanding training data to strengthen speech–text modal alignment \cite{liu2024improved}.

% \cmidrule(lr){3-7} \cmidrule(lr){8-9} \cmidrule(lr){10-12}
% & & Acc. ($\uparrow$) & Pre ($\uparrow$) & Rec ($\uparrow$) & F1 ($\uparrow$) & Pun ($\%$) 
% & Coa ($\uparrow$) & Fin ($\uparrow$) & Heg ($\uparrow$) & Hog ($\uparrow$) & Hop ($\uparrow$)\\
% \midrule

\section{Conclusion}
In this paper, we introduce APUN-Bench, an audio pun evaluation benchmark designed to comprehensively evaluate pun understanding in LALMs. We provide the first audio pun dataset, automate evaluation metrics and conduct extensive experiments on 10 open-source and proprietary LALMs, as well as 4 cascaded systems. Furthermore, by covering three stages: pun recognition, pun word location, and pun meaning inference, our benchmark systematically evaluates the model's capabilities across multiple dimensions of audio pun understanding. The experiment results show that while some models achieve promising performance on basic recognition tasks, they continue to face substantial challenges in pun word fine-grained location and inference, particularly for heterographic and homophonic puns. Through detailed analyses of positional biases and error types, we highlight the unique obstacles that audio puns pose and point to future directions for advancing more robust and human-like audio language understanding.

\section*{Limitations}
While this research offers new perspectives and insights into the understanding of audio puns, several limitations remain: (1) The scope of pun types extends beyond the three categories examined in this study, with more complex forms such as recursive and compound puns remaining outside its consideration. (2) Puns often arise in more complex speech environments, such as multi-turn dialogues, whereas our dataset is restricted to single-sentence instances. (3) The size of the real-world corpus still remains limited, which constrains the statistical robustness of our findings. However, despite these limitations, this paper represents the first systematic study of verbal puns, and its data and analyses offer valuable references for subsequent research on the intersection of language and speech understanding.
%尽管我们为语音双关语的理解提供了良好的视野和思路，然而仍然存在有一些限制。首先，双关语的类型并不局限于文章中提到的三种类型，还存在xxx等复杂类型，我们的文章并没有纳入对这些类型的考验。其次，双关语通常会存在多个复杂的语音环境下，例如对话场景等，我们目前仅考虑单一的双关语句子作为数据集。最后，真实数据的数量仍然有限。我们旨在未来的工作中进一步扩充数据集的广度

\section*{Ethical Considerations}
The Institutional Review Board (IRB) of our institution has approved the human studies presented in this paper. In addition, we have obtained permission to use data collection from the public pun website \textit{O. Henry Museum Pun-Off}.

\section*{Acknowledgments}
This research is supported by the Strong AI Lab and the Natural, Artificial, and Organisation Intelligence Institute at the University of Auckland. The first author of this research is funded by the China Scholarship Council (CSC). 

\bibliography{custom}

\appendix

\section{Baseline Models}
\label{Baseline Models}

\noindent\textbf{Qwen2-Audio-Instruct.}
\citet{Qwen2-Audio} present Qwen2-Audio, a large audio–language model that integrates an audio encoder with the Qwen LLM to process spoken queries, dialogues, and instruction-following tasks. The instruct variant evaluated here follows the official release, focusing on robust audio understanding across speech and environmental sound domains.

\noindent\textbf{SALMONN.}
\citet{tang2023salmonn} introduce SALMONN, which equips LLMs with general “hearing” abilities. It combines speech, audio event, and music encoders with a frozen LLM backbone, supporting tasks ranging from ASR and speech translation to audio captioning. SALMONN has been widely adopted as an open-source baseline for broad auditory competence.

\noindent\textbf{Audio-Reasoner.}
\citet{xie2025audioreasonerimprovingreasoningcapability} propose Audio-Reasoner, a reasoning-enhanced model trained with structured chain-of-thought supervision and a large corpus of annotated audio–text pairs (CoTA). It is designed to strengthen multi-step deduction and temporal reasoning, offering explicit reasoning traces alongside competitive performance.

\noindent\textbf{Qwen-2.5-Omni.}
\citet{Qwen2.5-Omni} extend the Qwen family with an omni-modal series capable of processing text, audio, and vision within a unified interface. The 3B and 7B variants are optimised for instruction following, open-domain spoken QA, and multimodal reasoning, providing strong baselines at different parameter scales.

\noindent\textbf{MiniCPM.}
\citet{yao2024minicpm} present MiniCPM-V, a lightweight multimodal model optimised for efficient deployment on mobile devices. Despite its compact scale, MiniCPM demonstrates performance comparable to GPT-4V on several multimodal tasks. Its audio-capable versions support real-time speech understanding for instruction-following scenarios.

\noindent\textbf{Audio Flamingo 3.}
\citet{goel2025audio} introduce Audio Flamingo 3 (AF3), a fully open state-of-the-art LALMs capable of multi-turn, multi-audio dialogue and long-context audio processing, enabling stronger cross-domain capabilities over speech, sound, and music. 

% \noindent\textbf{LLaSO-Base.}
% \citet{sun2024llaso} release LLaSO, an open framework for reproducible research in speech–language modelling. The 3.8B \emph{LLaSO-Base} model is trained exclusively on public corpora and provides a transparent reference point for large-scale audio–language evaluation.

\noindent\textbf{Omni-R1.}
\citet{rouditchenko2025omni} explore whether direct audio fine-tuning is necessary for competitive performance. Omni-R1 leverages large-scale instruction tuning on text paired with synthetic audio, achieving strong spoken understanding and reasoning without explicit raw-audio fine-tuning.

\noindent\textbf{MERaLiON-AudioLLM.}
\citet{li2024meralion} introduce MERaLiON-AudioLLM, a multilingual and multimodal large audio–language model. The family emphasises robustness to multilingual inputs, accents, and code-switching. In our evaluation, we reference the 10B variant (\emph{MERaLiON2}), which extends this line with improved large-scale instruction-following capabilities.

\noindent\textbf{GPT-4o-Audio.}
\citet{hurst2024gpt} document the GPT-4o family; we evaluate the audio-enabled version that accepts spoken input for transcription, dialogue, and multimodal reasoning. As a closed-source model, GPT-4o-Audio sets a strong commercial baseline for real-time, latency-sensitive applications.

\noindent\textbf{Gemini 2.0 Flash.}
The Gemini Flash models are proprietary systems released by Google DeepMind as part of the Gemini family \citet{jaech2024openai}. They are optimised for low-latency instruction following with speech input. We include Gemini 2.0 as a strong closed-source baseline to contextualise open-source performance.

\begin{table*}[t]
\resizebox{\textwidth}{!}{
\centering
\begin{tabular}{p{3cm}p{5cm}p{5cm}p{2cm}p{2cm}} % 第二列定义了宽度为8cm并自动换行
\toprule
Error Type & Definition & Example (Audio Transcription) & Prediction &Ground Truth \\ 
\midrule
\addlinespace[0.5ex]   % 给虚线下方加空隙
\multicolumn{5}{c}{\textit{Heterographic Puns \& Homophonic Puns}} \\ \hdashline
\addlinespace[0.5ex]   % 给虚线下方加空隙
Position Error (PE) & The model fails to provide the correct word corresponding to the given pun word and may instead return a word from another position in the sentence. & When asked to picture the perfect modern defensive weapon the Claymore springs to \textbf{mine}. & Claymore & mind \\
\addlinespace[0.5ex]   % 给虚线下方加空隙
\hdashline
\addlinespace[0.5ex]   % 给虚线下方加空隙
Pun Word Repeat (PWR) & The model outputs the given pun word itself rather than its intended meaning. & Did you hear about the nervous preacher ? He had sweaty \textbf{psalms}. & psalms & palm \\
\addlinespace[0.5ex]   % 给虚线下方加空隙
\hdashline
\addlinespace[0.5ex]   % 给虚线下方加空隙
Word Inference Error (WIE) (Similar) & The model infer an incorrect word, often manifested as another word with a similar pronunciation. & One ear of corn said to the other'You're getting \textbf{husky}'. & hoarse & husk\\
\addlinespace[0.5ex]   % 给虚线下方加空隙
\hdashline
\addlinespace[0.5ex]   % 给虚线下方加空隙
Word Inference Error (WIE) (No Similar) & The model infer an incorrect word, often manifested as another word with a completely different pronunciation. & George Westinghouse was a refrigerator \textbf{magnate}. & entrepreneur & magnet\\
\addlinespace[0.5ex]   % 给虚线下方加空隙
\hdashline
\addlinespace[0.5ex]   % 给虚线下方加空隙
Irrelevant Answer (IA) & The model’s answer is unrelated to the pun inference for the pun word. & She was only a Butcher's daughter, but there wasn't much more she could \textbf{loin}. & line human given an audio ... & learn\\
% \addlinespace[0.5ex]   % 给虚线下方加空隙
% \hdashline
% \addlinespace[0.5ex]   % 给虚线下方加空隙
% Phrase Omission \\
\midrule
\addlinespace[0.5ex]   % 给虚线下方加空隙
\multicolumn{5}{c}{\textit{Homographic Puns}} \\ \hdashline
\addlinespace[0.5ex]   % 给虚线下方加空隙
Reasoning Disorder (RD) & The model can only infer a single layer of meaning, where one pair of matches shows high similarity while the other pair fails to align. & How could I trust the ceiling fan installer when I knew he was always \textbf{screwing up}. & making mistakes AND causing trouble & making mistakes AND installing screws\\
\addlinespace[0.5ex]   % 给虚线下方加空隙
\hdashline
\addlinespace[0.5ex]   % 给虚线下方加空隙
Meaning Mismatch (MM) & The model generates double-meaning words or phrases that deviate from the intended meanings defined in the gold standard. & Can honeybee abuse lead to a \textbf{sting} operation? & steamboat AND steamship & undercover police activity AND painful insect bite\\
\addlinespace[0.5ex]   % 给虚线下方加空隙
\hdashline
\addlinespace[0.5ex]   % 给虚线下方加空隙
% Phrase Omission (PO) & The model either refuses to generate an inference or produces only a single interpretation. & & \\ %模型在推理阶段遗漏了期望输出，即拒绝输出或者只输出单个含义。\\
% \addlinespace[0.5ex]   % 给虚线下方加空隙
% \hdashline
% \addlinespace[0.5ex]   % 给虚线下方加空隙
Irrelevant Answer (IA) & The model’s answer is irrelevant style of inference meaning or incorrect styles for the pun word. & Then there was the occasion I spotted a health - oriented cafe displaying a sign reading' California \textbf{Shakes}.' Obviously, I thought. & The two meanings of 'shake' in this context & earthquake AND milkshake\\
\bottomrule
\end{tabular}
}
\caption{Error types of pun inference in heterographic, homophonic and homographic puns}
\label{pun_inference_error}
\end{table*}

\section{Synthesis Process}
\label{Synthesis Process}
We use Parler-TTS for speech synthesis. This open-source, lightweight text-to-speech system is capable of generating high-quality, natural-sounding speech conditioned on various speaker characteristics, such as gender, pitch, and speaking style \citep{lacombe-etal-2024-parler-tts}. To ensure diversity, we utilize the 34 default speaker embeddings and randomly assign one to each generated utterance, as shown in Table \ref{tab:speakers}. Finally, a manual filtering step is conducted to screen the missing words or unclear pronunciations to improve the overall quality of the speech data.

\begin{table}[t]
\centering
\small
\begin{tabular}{llll}
\toprule
Laura   & Gary    & Jon     & Lea \\
Karen   & Rick    & Brenda  & David \\
Eileen  & Jordan  & Mike    & Yann \\
Joy     & James   & Eric    & Lauren \\
Rose    & Will    & Jason   & Aaron \\
Naomie  & Alisa   & Patrick & Jerry \\
Tina    & Jenna   & Bill    & Tom \\
Carol   & Barbara & Rebecca & Anna \\
Bruce   & Emily   &         &      \\
\bottomrule
\end{tabular}
\caption{List of the 34 speakers used in Parler-TTS.}
\label{tab:speakers}
\end{table}
%我们使用了Parler-TTS俩合成语音，这是一个开源的轻量级的文本转语音的模型，能够根据不同说话者来生成高质量、自然的语音（例如性别，音调，风格），为了确保获得语音的多样性，我们使用了默认34位说话者的音素并在生成每一条语音文件的时候随机指定一位说话者。最后，执行手动过滤步骤，对于遗漏单词或者发音不清的case执行重复操作以达到提高合成语音数据的质量。

% \subsection{Similarity Threshold}
% \label{Threshold}

\section{Supplementary Analysis}

\subsection{Error Types}
We summarize the error types of LALMs in the audio pun inference stage, as presented in Table \ref{pun_inference_error}.

\textbf{Position Error (PE).} The model fails to provide the correct word corresponding to the given pun word and may instead return a word from another position in the sentence.

\textbf{Pun Word Repeat (PWR).} The model outputs the given pun word itself rather than its intended meaning.

\textbf{Word Inference Error with Similar (WIE).} The model infer an incorrect word, often manifested as an other word with a similar pronunciation.

\textbf{Word Inference Error with No Similar (WIE).} The model infer an incorrect word, often manifested as another word with a completely different pronunciation.

\textbf{Irrelevant Answer (IA).} The model can only infer a single layer of meaning, where one pair of matches shows high similarity while the other pair fails to align.

\textbf{Reasoning Disorder (RD).} The model generates double meaning words or phrases that deviate from the intended meanings defined in the gold standard.

\textbf{Meaning Mismatch (MM).} The model’s answer is irrelevant style of inference meaning or incorrect styles for the pun word.

% \section{Data Analysis}
% \label{data analysis}
\subsection{Are LALMs Biased Toward Certain Types of Pun Word Location?}

In the pun location stage, different models show notable performance gaps between coarse and fine-grained locations, with differences of around 20\% being common across models. While this indicates that the models are generally consistent in transcribing and recognizing audio puns, we further explore their recognition capabilities across specific pun categories. Table \ref{pun-location} presents the performance of various models on three categories of puns, evaluated through both coarse-grained and fine-grained tasks in synthetic dataset. The results reveal that most models perform significantly better on homographic puns, excelling in both location detection and exact matching. For heterographic and homophonic puns, although their accuracy in coarse location is comparable, the accuracy of fine-grained location for homophonic puns is substantially lower. This finding highlights that existing models still encounter considerable difficulties in accurately reproducing fully homophonic puns.
\begin{table}[t]
\resizebox{\columnwidth}{!}{
\centering
\begin{tabular}{lcccccc}
\toprule
\multirow{2}{*}{\textbf{Model}} & \multicolumn{2}{c}{\textbf{Heterographic}} & \multicolumn{2}{c}{\textbf{Homophonic}} & \multicolumn{2}{c}{\textbf{Homographic}} \\
\cmidrule(lr){2-3} \cmidrule(lr){4-5} \cmidrule(lr){6-7}
& Coa ($\uparrow$) & Fin ($\uparrow$) & Coa ($\uparrow$) & Fin ($\uparrow$) & Coa ($\uparrow$) & Fin ($\uparrow$)\\
\midrule
SALMONN &32.37 &17.12 &35.86 &8.55 & 35.32 & 28.81\\
Qwen2-Audio-Instruct  &52.14 &29.47 &44.41 &13.81 & 51.51 & 37.41\\
Audio-Reasoner  &50.50 &29.34 &39.47 &11.18 & 52.90 & 39.11\\
Qwen-2.5-omni      & 53.53 &36.39 &56.91 &21.38 & 66.54 & 61.42\\
MiniCPM  & 56.30 &34.13 & 62.50 &20.06 & 61.58 & 56.85\\
Omni-R1  & 58.06 &38.16 &60.53 &21.71 & 67.54 & 62.89\\
MERaLiON2 & 55.16 & 31.36 & 62.50 & 17.10 & 63.21 & 57.70 \\
Gemini 2.0 Flash \ding{79} &60.96 &36.90 & 70.72 &35.19 & 67.62 & 58.48 \\
GPT4o-Audio \ding{79} &\textbf{74.06} &\textbf{44.20} &\textbf{76.64} &\textbf{42.43} & \textbf{76.99} & \textbf{67.62}\\
\bottomrule
\end{tabular}
}
\caption{Evaluation results of the Synthetic Dataset on APUN-Bench. We evaluate various LALMs on our proposed APUN-Bench. \ding{79} represents the proprietary LALMs.}
\label{pun-location}
\end{table}

\subsection{Part-of-Speech Distributions}
Additionally, we analyze the part-of-speech distributions of pun words in both the synthetic and real-world subsets using SpaCy, identifying the top five most frequent categories as shown in Table \ref{Part-of-Speech Distributions}. The findings indicate that nouns and verbs are more likely to serve as pun-bearing words, implying that certain lexical categories are inherently more productive in pun construction. This observation provides insight into the lexical bias underpinning pun formation.
\begin{table}[t]
\resizebox{\columnwidth}{!}{
\centering
\begin{tabular}{lccccc}
\toprule
\textbf{Model} & \textbf{NOUN} & \textbf{ADV} & \textbf{VERB} & \textbf{ADJ} & \textbf{PROPN}\\
\midrule
Synthetic Dataset &954	&312	&246	&202&	146\\
Real-world Dataset  &208&	36	&131	&81&	93 \\
\bottomrule
\end{tabular}
}
\caption{The part-of-speech distributions between Synthetic and Real-world Datasets.}
\label{Part-of-Speech Distributions}
\end{table}

\subsection{Pun Transcript vs Full Audio?}
% 为了清晰表明语音双关语中的音频信号对于端到端模型的影响，我们在Real-world数据集上对纯文本和纯音频的输入进行了对比，结果如表X所示，总体而言，在双关语识别和词语定位任务中，文本模型的表现均好过语音模型，而在推理任务中，开源文本模型的推理能力要比语音模型要弱。这表明，声学信息的加入会促使规模较小的模型更好地处理双关语词推理
To clearly evaluate the respective contributions of audio signals and textual content in audio pun understanding, we conduct a controlled comparison between text-only and audio-only inputs on real-world datasets, with results reported in Table \ref{text vs audio}. For both the pun recognition and pun word location tasks, text-based models consistently outperform speech-based models, which can be attribute to the limited robustness of current ASR systems when transcribing speech puns. 

In contrast, for the pun meaning inference task, open-source text-only models exhibit inferior reasoning performance compared to their audio-based counterparts. This finding suggests that, under the assumption of perfect or near-perfect textual input, performance gains observed in recognition-oriented tasks may primarily reflect robustness to transcription quality. However, for inference tasks that require resolving ambiguities intrinsic to spoken puns, acoustic information provides complementary evidence that benefits end-to-end speech models, particularly those with limited model capacity.
% We compared the performance on the Real-world dataset with plain text and audio inputs across Qwen2.5, GPT, and the results are shown in Table \ref{text vs audio}. Overall, in terms of performance with plain text, it is comparable to the conclusions of \citep{xu2024good}.

\subsection{Word Error Rate}
%我们分别报告了SenseVoiceSmall和Whisper-Large在合成数据和真实数据上的词错误率，如表x所示。整体而言，真实数据的词错误率要比合成数据更高，表明真实数据中的环境和人为发音等因素会干扰音频转录的准确率，是可能导致真实数据。在双关语类别上，homographic的词错误率整体要比heterographic和homophonic要低，这说明转录模型在识别具有音频歧义的词语上仍具有提升空间。
%基于级联模型，我们统计了
\label{Word Error Rate section}
We report the word error rate (WER) of the SenseVoiceSmall and Whisper-Large ASR models on both synthetic and real data to quantify their transcription errors on audio puns. The results are shown in Table~\ref{Word Error Rate}. Overall, the WER on real-world data is higher than that on synthetic data, indicating that more complex environmental noise, speaker variability, and natural pronunciation variations in real speech interfere with the accuracy of audio transcription. Further analysis from the perspective of pun type reveals that, under both data settings, the WER for homographic puns is consistently lower than that for heterographic and homophonic puns. Compared to phonetic ambiguity caused by near-homophones, the result indicates that ASR models are more robust in handling words that involve only semantic ambiguity without introducing substantial phoneme confusion; in contrast, for puns with pronounced audio-level ambiguity (such as heterographic and homophonic puns), existing transcription models still have considerable room for improvement.

\begin{table}[t]
\resizebox{\columnwidth}{!}{
\centering
\begin{tabular}{ccccccc}
\toprule
\multirow{2}{*}{\textbf{ASR Model}} & \multicolumn{3}{c}{\textbf{Synthetic}} & \multicolumn{3}{c}{\textbf{Real-world}} \\
 & Heg ($\downarrow$) & Hog ($\downarrow$) & Hop ($\downarrow$) & Heg ($\downarrow$) & Hog ($\downarrow$) & Hop ($\downarrow$)\\
\midrule
SenseVoiceSmall &0.144	&0.126 & 0.151 & 0.177 & 0.134 & 0.269\\
Whisper-Large  &0.111 &	0.097 & 0.123 & 0.154 & 0.093 & 0.221	\\
\bottomrule
\end{tabular}
}
\caption{Word Error Rate (WER) of SenseVoiceSmall and Whisper-Large on synthetic and real-world datasets.}
\label{Word Error Rate}
\end{table}

\subsection{Time Consumption Analysis}
\label{Time Consuming Section}
%我们统计了级联模型和端到端语音模型在处理双关语任务时所需的时间消耗。为了防止网络等额外因素的影响，我们只对本地开源LALMs和级联模型在真实世界数据集上进行实验，实验设置采用了一块A100 80G的GPU，并选择使用MiniCPM和Qwen-2.5-omni作为LALMs，以及四种级联系统的组合。实验结果表明，相较于端到端语音模型，级联模型在处理双关语任务时需要更长的推理时间，这一现象主要源于级联系统需要顺序执行语音转写和文本处理两个阶段，从而引入额外的计算开销。相比之下，LALMs 通过端到端建模语音到语义的映射，显著降低了整体推理延迟。这一结果表明，采用 LALMs 进行语音双关语理解在效率上具有明显优势。

We statistically analyze the time consumption of cascaded systems and end-to-end speech models in audio pun dataset, as shown in Table \ref{Time Consuming}. To prevent the influence of additional factors such as network latency, we only conduct experiments on local open-source LALMs and cascaded systems on real-world datasets. The experimental setup used an A100-SXM4-80GB, and we select MiniCPM and Qwen-2.5-omni as LALMs, along with a combination of four cascaded systems. Experimental results show that compared to end-to-end speech models, cascaded systems require longer inference time when processing puns. This is mainly due to the fact that cascaded systems need to sequentially execute the speech transcription and text processing stages, introducing additional computational overhead. In contrast, LALMs reduce the overall inference latency by modeling the speech-to-semantic mapping end-to-end. This result demonstrates that using LALMs for speech pun understanding has a clear efficiency advantage.
\begin{table}[t]
\resizebox{\columnwidth}{!}{
\centering
\begin{tabular}{lcccc}
\toprule
\textbf{Model} & \textbf{Architecture} & \textbf{ASR (s)} & \textbf{Reasoning (s)} & \textbf{Total (s)}\\
\midrule
MiniCPM & End-to-End & - & 0.61 & 0.61\\
Qwen-2.5-omni & End-to-End & - &0.91 & 0.91\\
\addlinespace[0.5ex]   % 给虚线下方加空隙
\hdashline
\addlinespace[0.5ex]   % 给虚线下方加空隙
SenseVoiceSmall+MiniCPM &Cascaded	& 0.42	&0.54	&0.74\\
Whisper-Large+MiniCPM  &Cascaded&	1.31	&0.54	&1.85 \\
SenseVoiceSmall+Qwen-2.5-Instruct &Cascaded	& 0.42	&0.34&0.76\\
Whisper-Large+Qwen-2.5-Instruct  &Cascaded&	1.31	&0.34&1.65 \\
\bottomrule
\end{tabular}
}
\caption{Time consumption analysis in real-world dataset between open-source LALMs and cascaded systems.}
\label{Time Consuming}
\end{table}

\begin{table*}[t]
\centering
\resizebox{2\columnwidth}{!}{
\begin{tabular}{lcccccccccc}
\toprule
\multirow{2}{*}{\textbf{Model}} & 
\multirow{2}{*}{\textbf{Size}} & 
\multicolumn{5}{c}{\textbf{Pun Recognition}} & 
\multirow{2}{*}{\textbf{Pun Location}} & 
\multicolumn{3}{c}{\textbf{Pun Inference}} \\
\cmidrule(lr){3-7}  \cmidrule(lr){9-11}
& & Acc. ($\uparrow$) & Pre ($\uparrow$) & Rec ($\uparrow$) & F1 ($\uparrow$) & Pun(\%) & 
& Heg ($\uparrow$) & Hog ($\uparrow$) & Hop ($\uparrow$) \\
\midrule
\addlinespace[0.5ex]   % 给虚线下方加空隙
\multicolumn{11}{c}{\textit{Audio}} \\ \hdashline
\addlinespace[0.8ex]   % 给虚线下方加空隙
MiniCPM & 8.7B & 74.35 & 98.08 & 61.62 & 75.68 & 40.80 & 38.03 & 43.69 & 34.86 & 54.79 \\
Qwen-2.5-omni & 7B & 71.68 & \textbf{99.09} & 56.79 & 72.21 & 37.12 & 42.85 & 50.00 & 35.78 & 57.53 \\
Gemini 2.0 Flash \ding{79} & -&	87.51&	84.82	&98.51	&91.15&	60.98&	51.03&	62.78&	47.36&	82.85\\
GPT4o-Audio \ding{79} & -&85.36&	87.02&	92.56	&89.71	&66.89	&50.08	&60.98	&50.70&	76.71 \\
\midrule
\addlinespace[0.5ex]   % 给虚线下方加空隙
\multicolumn{11}{c}{\textit{Text}} \\ \hdashline
\addlinespace[0.8ex]   % 给虚线下方加空隙
MiniCPM & 8.7B & 65.44 & 100 & 46.64 & 63.61 & 30.21 & 78.14 & 34.08 & 34.38 & 45.20\\
Qwen-2.5-Instruct & 7B & 83.61 & 99.09 & 75.38 & 85.63 & 49.28 & 76.93& 39.01 & 39.73 & 43.83\\
Gemini 2.0 Flash \ding{79} & -& 91.60&	98.46	&88.46&	93.20&	58.19&	85.22&	72.19&	60.00	&93.15\\
GPT4o \ding{79} & -& 95.08&	100&	92.42&	96.06& 59.87&	93.03&	74.88&	69.47&	93.15 \\
\bottomrule
\end{tabular}
}
\caption{Comparison between text-only and audio models in real-world dataset. \ding{79} represents the proprietary LALMs and the symbol '–' indicates that the model size is not publicly disclosed.}
\label{text vs audio}
\end{table*}

\section{Annotation Prompts}
\label{Annotation Prompts}
We use the following prompt to instruct Claude-Opus (Claude-Opus-4-1) to identify and locate puns in transcribed speech. The prompt includes a clear task description, output specification, and a few manually annotated examples with corrective feedback to improve the accuracy of the auxiliary annotation. Some details are shown as Figure \ref{pun_location}.

% \section{Dataset Link}
% An anonymized subset of the dataset is available at: \hyperlink{https://anonymous.4open.science/r/audio-pun-public-5046}{https://anonymous.4open.science/r/audio-pun-public-5046}

\section{Human Involvement}

\subsection{Annotation}
\label{human annotation}
%在真实数据的获取方面，我们招募了10个具有良好英语背景的参与者进行录制，平均每一个人录制30-60条不定数量的双关语句子，我们给予每条录制的音频报酬为2元。
For real-world data recordings, we recruit 10 adult anonymous participants with strong English-speaking backgrounds, all of whom hold a college education and represent diverse majors from different countries. To ensure diversity in speech, each participant records approximately 30–60 pun sentences, and we provide a compensation of \$0.3 per sentence recording.

\subsection{Human Verification}
\label{Human Verification}
Human verification is carried out by two experts, each with a master’s or doctoral degree and strong English proficiency. The review process is divided into two parts, including both synthetic dataset and real-world dataset:

\textbf{Audio Quality:} Experts examine the audio files to identify any missing or under-recorded segments and to ensure that the recordings are clear and intelligible. Each audio clip averages no more than 8 seconds in length. To preserve the authenticity of natural speech, accent variation is not restricted. For quality assessment, we randomly sample 300 audio files from the dataset. Approximately 1.5\% of the samples require re-recording due to defects, yielding a low overall defect rate and confirming the high quality of the audio data.

\textbf{Annotation Verification:} After the initial annotation stage, we conduct a systematic quality check. Experts confirm the presence of puns, specify alternatives for heterographic and homophonic puns, and validate the accuracy of dual meanings in homographic puns. A random sample of 300 data points is selected for cross-validation. On average, reviewers spend 45–60 seconds evaluating each speech sample. Each entry is independently verified by each experts, and inter-annotator agreement is measured using \textit{Cohen’s K}, which reaches 0.92, indicating substantial consistency across annotators.
% \begin{itemize}
%     \item \textbf{Audio quality:} Annotators are asked to verify whether the audio files contain any missing or under-recorded segments and ensure that the recordings are clear and intelligible. The average duration of each audio clip does not exceed 10 seconds. To preserve the authenticity of natural speech, accent variation is not limited.  

%     \item \textbf{Annotation verification:} This involves identifying puns, marking word positions, specifying alternatives for heterographic and homophonic puns, and annotating dual meanings for homographic puns. Each entry was independently verified by two annotators, with a third annotator resolving disagreements when necessary.  

% \end{itemize}
% On average, reviewers spent 45–60 seconds reviewing each speech sample. To assess annotation consistency, 300 data points were randomly selected from the dataset for cross-validation.

% \subsection{Human Evaluation}

\begin{table*}[t]
\centering
\resizebox{1.4\columnwidth}{!}{
\begin{tabular}{lccccccc}
\toprule
\multirow{2}{*}{\textbf{Model}} & \multicolumn{2}{c}{\textbf{Pun Recognition}} & \multicolumn{2}{c}{\textbf{Pun Location}} & \multicolumn{3}{c}{\textbf{Pun Inference}} \\
\cmidrule(lr){2-3} \cmidrule(lr){4-5} \cmidrule(lr){6-8}
& Acc. ($\uparrow$) & Pun ($\%$) & Coa ($\uparrow$) & Fin ($\uparrow$) & Heg ($\uparrow$) & Hog ($\uparrow$) & Hop ($\uparrow$)\\
\midrule
SALMONN & 63.33 & 50.00 & 43.33 & 23.33 & 18.33 & 28.33 & 18.33\\
Qwen2-Audio-Instruct  & 51.66 & 43.33 & 58.33 & 33.33 & 30.00 & 23.63 & 28.33\\
Audio-Reasoner  & 58.33 & 50.00 & 63.33 & 40.00 & 31.67 & 34.61 & 35.42\\
Qwen-2.5-omni      & 78.33 & 58.33 & 60.00 & 36.67 & 43.33 & 31.66 & 65.00 \\
MiniCPM  & 53.33 & 33.33 & 53.33 & 30.00 & 46.67 & 44.06 & 58.33 \\
Omni-R1  & 75.00 & 65.00 & 56.67 & 33.33 & 45.00 & 44.82 & 63.89 \\
MERaLiON2 & 73.33 & \textbf{86.67} & 56.67 & 26.67 & 51.67 & 36.66 & 65.00\\
Gemini 2.0 Flash \ding{79} & \textbf{81.67} & 63.33 & 70.00 & 41.66 & 51.67 & 53.33 & 81.67\\
GPT4o-Audio \ding{79} & 73.33 & 86.67 & 73.33 & 46.67 & 53.33 & 52.54 & 78.33 \\
\hdashline
\addlinespace[0.5ex]   % 给虚线下方加空隙
Human & 76.67 & 41.67 & \textbf{76.67} & \textbf{60.00} & \textbf{78.33} & \textbf{73.33} & \textbf{90.00} \\
\bottomrule
\end{tabular}
}
\caption{Comparison with Human Evaluation and LALMs. \ding{79} represents the proprietary LALMs.}
\label{human evaluation}
\end{table*}

\subsection{Human Evaluation}
We randomly sample 60 data points for manual evaluation from each stage to compare the performance of current mainstream LALMs with that of human participants. Task-specific questionnaires are designed and distributed to three anonymous volunteers via the scientific research platform Prolific \footnote{https://www.prolific.com/}, targeting primarily native English speakers from the UK and the USA. Each participant receives a reward of \$5.4 per completed questionnaire. Table \ref{human evaluation} presents the evaluation results, indicating that while some state-of-the-art LALMs achieve performance comparable to human evaluators in certain pun understanding tasks, they still lag significantly behind humans in others, such as pun inference. In addition, we observe that human evaluators exhibit a conservative bias when identifying puns, tending to classify utterances as non-puns. However, it is worth noting that pun understanding relies on rich linguistic and world knowledge, representing a high-level language comprehension task. Therefore, this task calls for larger-scale and more diverse human evaluations, which we plan to pursue in future work.

\begin{figure*}[htbp]
    \centering
    \begin{tcolorbox}
    /* Definition */

    You are an expert linguist with specialized knowledge of pun research. Your role is to annotate transcribed spoken sentences by identifying the pun word. A pun word is a word or phrase that exploits multiple meanings, sound similarities, or semantic ambiguities to create humor or wordplay. \\

    /* Instruction */
    
    For each given sentence, identify the pun word. 
If there is no pun word, output "None". 
Only output the pun word (or the minimal pun phrase), without explanations or additional text. 
Be consistent with the format shown in the examples. \\

    /* Examples */
    
    Pun Sentence: Dentists don't like a hard day at the orifice.
    
    Output: orifice

    Pun Sentence: A discussion of digging a new mine shaft was too deep for him.
    
    Output: deep 

    ... \\
    
    /* Test Data */
    
    Pun Sentence: He didn't tell his mother that he ate some glue. His lips were sealed.
    
    Output:
    
    \end{tcolorbox}
    \caption{Prompts used to guide Claude-Opus-4-1 in pun location for auxiliary annotation.}
    \label{fig:pun_location}
\end{figure*}

\begin{figure*}[htbp]
    \centering
    \begin{tcolorbox}
    /* Definition */

    You are an expert linguist with specialized knowledge of pun research.  
    Your role is to annotate transcribed spoken sentences by inferring the alternative word 
    that the pun word replaces or plays upon.  
    A pun word is a word or phrase that exploits multiple meanings, sound similarities, 
    or semantic ambiguities to create humor or wordplay. \\ 
    
    /* Instruction */
    
    For each given pun sentence, infer the intended alternative word.  
    Only output the alternative word (or minimal phrase) without explanations or additional text.  
    Ensure consistency with the format shown in the examples. \\

    /* Examples */
    
    Pun Sentence: Dentists don't like a hard day at the orifice.

    Pun word: orifice
    
    Output: office

    ... \\
    
    /* Test Data */
    
    Pun Sentence: Geologists can be sedimental about their work.

    Output: sedimental
    
    Output: 
    
    \end{tcolorbox}
    \caption{Prompts used to guide Claude-Opus-4-1 in pun inference for auxiliary annotation (for heterographic and homophonic puns).}
    \label{fig:inference_hete}
\end{figure*}

\begin{figure*}[htbp]
    \centering
    \begin{tcolorbox}
    /* Definition */
    
    You are an expert linguist with specialized knowledge of pun research.  
Your role is to annotate transcribed spoken sentences by inferring the two distinct meanings 
of a homographic pun word.  A pun word is a word or phrase that exploits multiple meanings, sound similarities, or semantic ambiguities. \\

    /* Instruction */
    
    For each given pun sentence, identify the pun word and infer its two distinct meanings. Output exactly two replacement words or phrases, each representing one meaning of the pun word. Provide only the two items, separated by a comma, with no additional text.  
Ensure consistency with the format shown in the examples.   \\

    /* Examples */
    
    Pun Sentence: A discussion of digging a new mine shaft was too deep for him

    Pun Word: deep
    
    Output: profound, physically deep

    ... \\
    
    /* Test Data */
    
    Pun Sentence: If at first you don't succeed then skydiving is not for you.

    Pun Word: succeed
    
    Output:
    
    \end{tcolorbox}
    \caption{Prompts used to guide Claude-Opus-4-1 in pun inference for auxiliary annotation (for homographic puns).}
    \label{fig:inference_homo}
\end{figure*}

\end{document}